\documentclass[a4paper,amsmath,amssymb,superscriptaddress,aps,prb,reprint,showpacs]{revtex4-2}
\usepackage{xcolor,graphicx}
\usepackage{subfigure}
\usepackage{MnSymbol}
\usepackage{braket}
\usepackage{tabularx}
\usepackage{booktabs}
\usepackage{tikz}
\usepackage{pgfplots}
\usepackage[utf8]{inputenc}
\usepackage{floatrow}
\newfloatcommand{capbtabbox}{table}[][\FBwidth]

\renewcommand{\vec}[1]{\mathbf{#1}}

\begin{document}

\title{Comparing the influence of Floquet dynamics in various Kitaev-Heisenberg materials}

\author{Pascal Strobel}

\affiliation{%
Institut f\"ur Funktionelle Materie und Quantentechnologien,
Universit\"at Stuttgart,
70550 Stuttgart,
Germany}

\author{Maria Daghofer}
\affiliation{%
Institut f\"ur Funktionelle Materie und Quantentechnologien,
Universit\"at Stuttgart,
70550 Stuttgart,
Germany}
\affiliation{Center for Integrated Quantum Science and Technology, University of Stuttgart,
Pfaffenwaldring 57, 
70550 Stuttgart, Germany}

\date{\today}

\begin{abstract}
In this paper we examine the possibility of Floquet engineering in the
three candidate Kitaev materials $\mathrm{Na}_2\mathrm{IrO}_3$,
$\alpha$-$\mathrm{Li}_2\mathrm{IrO}_3$ and
$\alpha$-$\mathrm{RuCl}_3$. We derive an effective Floquet Hamiltonian
and give an approximation for heating
processes arising from doublon holon propagation.This suggests that compounds with stronger
  Hund's-rule coupling are less prone to heating. We then 
investigate the impact of light frequency and amplitude on 
magnetic interaction terms up to third-nearest--neighbor and find that
third-neighbor Heisenberg coupling $J_3$ is very susceptible to tuning
by circularly polarized light. Finally, we discuss uses of  linear polarized light in selectively tuning single bond directions. 
\end{abstract}

\maketitle
\section{Introduction}\label{intro}
Light-matter interactions with periodic driving have gathered 
significant attention recently
\cite{RevModPhys.89.011004,Oka,Mentink_2017}. In these systems the Floquet theorem is applicable,
and they can be described with an effective time-independent Floquet Hamiltonian
\cite{Mentink_2015,Claassen2017,PhysRevLett.116.125301,PhysRevLett.121.107201}. Frequency,
amplitude, and polarization are then capable of changing intrinsic
interactions of the system. This paves the way  
to tune magnetic properties of materials via the driving field. While
for large time scales the system thermalizes, the Floquet Hamiltonian
has been argued to be valid at short, but experimentally accessible,
timescales  \citep{KUWAHARA201696}.  

Floquet engineering is especially promising for candidate
Kitaev-Heisenberg materials \cite{Winter_2017}.  
Those materials are supposed to host a Kitaev spin liquid (KSL) for an
idealized structure. However in reality most of them show a zig-zag
ordering
\cite{PhysRevB.83.220403,PhysRevB.85.180403,PhysRevLett.108.127204,PhysRevB.92.235119}. Tuning
these materials from zig-zag to the KSL has been a focal point
of theoretical and experimental research. 
While tuning via pressure
\cite{PhysRevB.96.205147,PhysRevB.97.245149,PhysRevB.103.L140402},
magnetic field
\cite{PhysRevLett.118.187203,PhysRevLett.119.037201,PhysRevB.95.180411,PhysRevLett.119.227208,PhysRevLett.120.117204},
graphen substrates \cite{PhysRevLett.123.237201} and chemical doping
\cite{PhysRevMaterials.1.052001,PhysRevB.99.214410,PhysRevB.102.094407}
has been investigated extensively, the  approach of driving the
Kitaev phase via a periodic light field is yet relatively unexplored.   

Driving with light periodic in time might allow to alter the
relative strength of the interactions, and
therefore give rise to a KSL. In particular, materials that are already
proximate to the Kitaev phase should be a good starting
point. Such materials are, e.g., iridates like
\(\mathrm{Na}_2\mathrm{IrO}_3\) \cite{PhysRevB.88.035107} and
$\alpha$-$\mathrm{Li}_2\mathrm{IrO}_3$ \cite{PhysRevLett.113.197201}
and the ruthenate $\alpha$-$\mathrm{RuCL}_3$
\cite{PhysRevB.91.241110}. Since these materials have already been studied
in absence of a light field
\cite{PhysRevB.90.205116,PhysRevLett.113.197201,PhysRevLett.108.127204,PhysRevB.92.235119,PhysRevB.93.214431}, we base our studies
on already well-established models~\cite{PhysRevB.88.035107}. 
Previous studies on Floquet engineering \cite{PhysRevB.103.L100408}
have focused on tuning nearest--neighbor interactions in
$\alpha$-$\mathrm{RuCl}_3$ with circularly polarized light and highlighted the possibility of
tuning Heisenberg interactions. However
this phenomenon was just found after changing signs of a specific parameter. For
iridates studies on Floquet engineering are still lacking. 

In this paper we study the capabilities of Floquet engineering, for
$\alpha$-$\mathrm{RuCl}_3$ as well as for
$\alpha$-$\mathrm{Li}_2\mathrm{IrO}_3$ and
$\mathrm{Na}_2\mathrm{IrO}_3$.
First we consider an approximate
approach to capture heating around the
resonances~\cite{PhysRevB.99.205111} in order to identify frequencies
where it is small.  This gives an insight on which of these materials are
well suited for Floquet engineering.
Introducing linear polarized light instead of circular polarized light, interactions become bond
dependent. This gives the possibility to tune the interactions via the
light angle. We finally discuss third-nearest--neighbor
Heisenberg coupling.  It was found to be significant in ruthenates (and iridates) and is
argued to be responsible for the zig-zag magnetic
ordering~\cite{PhysRevB.93.214431}.  

The paper is structured as follows: In Sec.~\ref{theory} we derive
the effective Floquet Hamiltonian for the Kitaev-Heisenberg model,
applying the results of
\cite{PhysRevB.99.205111,PhysRevLett.121.107201,PhysRev.138.B979} to
the already well established Kitaev-Heisenberg
model~\cite{PhysRevLett.102.017205,PhysRevLett.112.077204}. We
introduce the off-resonance limit of the model as well as a
possibility to describe the system near resonances.  
In Sec.~\ref{res} we discuss the results for the 
materials $\alpha$-$\mathrm{RuCL}_3$,
$\alpha$-$\mathrm{Li}_2\mathrm{IrO}_3$ and
$\mathrm{Na}_2\mathrm{IrO}_3$. In Sec.~\ref{heat} we study heating,
and determine frequencies  where Floquet
driving without any considerable heating is possible. For these
frequencies we then investigate the effects of Floquet driving on the
Kitaev, Heisenberg and $\Gamma$ terms in Sec.~\ref{eng}. 
Furthermore we also consider third-nearest--neighbor Heisenberg terms (Sec.~\ref{tnn}) and  the impact of the polarization angle on the bond
dependent Kitaev, Heisenberg and $\Gamma$ interactions (Sec.~\ref{angle}).   
Section~\ref{summary} concludes the paper and gives a summary of the results
obtained as well as an outlook to promising future studies.

\section{Theory}\label{theory}

\subsection{Effective model}\label{eff}
The materials considered in this paper have strong Coulomb repulsion,
stabilizing  a $d^5$ configuration with one hole residing on each
site. 
Taking into account virtual excitations with two holes on one site via second order perturbation
theory, gives rise to a Kugel-Khomskii--type model
\cite{Kugel__1982}. Due to strong spin-orbit coupling (SOC) present in
the considered materials, we can simplify the Hamiltonian further
with a projection into the $j_{\mathrm{eff}}=1/2$ pseudospin basis. This leads to the well-known Kitaev-Heisenberg model
\cite{PhysRevLett.105.027204}. 
 
Coupling to a periodic light field
enters via Peierls substitution into the Hubbard model 
\begin{align}
H_{\mathrm{kin}}(t)=&-\sum_{\substack{\gamma,\sigma\\\braket{ij}_{\gamma}}}\underbrace{\vphantom{\mathbf{c}_{i,\sigma}^{\dagger}\mathbf{T}^{\gamma}\mathbf{c}_{i,\sigma}}e^{i(\vec{r}_j-\vec{r}_i)\vec{A}(t)}}_{\Omega
  (t)}\underbrace{\mathbf{c}_{i,\sigma}^{\dagger}\mathbf{T}^{\gamma}\mathbf{c}_{i,\sigma}}_{\hat{v}_{ij}}, 
\label{eq:EQ1}
\end{align}
where $\mathbf{c}_{i}=(c_{i,x,\sigma},c_{i,y,\sigma},c_{i,z,\sigma})$
$\left(\mathbf{c}_{i}^{\dagger}\right)$ annihilates (creates) a hole at site $i$ with spin $\sigma$.
For $d$-orbitals, we used the notation $yz\,(x)$, $zx\,(y)$, and $xy\,(z)$ introduced in \cite{PhysRevLett.112.077204}. The bond-dependent hopping matrices $\mathbf{T}^{\gamma}$ are given by
\begin{align}
\mathbf{T}^{z}=\begin{pmatrix}
t^{z}_1 & t^{z}_2 &t^{z}_4\\
t^{z}_2&t^{z}_1&t^{z}_4\\
t^{z}_4&t^{z}_4&t^{z}_3
\end{pmatrix},\,
\mathbf{T}^{x}=\begin{pmatrix}
t_3^{x} & t_4^{x} &t_4^{x}\\
t_4^{x}&t_1^{x}&t_2^{x}\\
t_4^{x}&t_2^{x}&t_1^{x}
\end{pmatrix}
,\, \mathbf{T}^{y}=\begin{pmatrix}
t_1^{y} & t_4^{y} &t_2^{y}\\
t_4^{y}&t_3^{y}&t_4^{y}\\
t_2^{y}&t_4^{y}&t_1^{y}
\end{pmatrix}.
\label{eq:EQ2}
\end{align}
 
Light interaction $\Omega (t)$ depends on the vector potential
$\mathbf{A}(t)$ and the vector $\mathbf{r}_{ij}$, between nearest--neighbors (NN) 
 $i$ and $j$.  In this paper we will primarily focus on linear
polarized (LP) light, i.e. the vector potential $\mathbf{A}(t)=(E_x,E_y)\sin(\omega t)$. With this at hand $\Omega(t)$ becomes 
$\sum_{l=-\infty}^{\infty}\mathcal{J}_{-l}(u_{ij})e^{-i\,l\omega  t},$
where $\mathcal{J}_{-l}(u_{ij})$ are Bessel functions of first kind and
$u_{ij}^{\gamma}=e/\omega\,\mathbf{r}^{\gamma}_{ij}E_0(\cos(\varphi),\sin(\varphi))$.
The on site interactions can be described with a Kanamori Hamiltonian \cite{Kanamori}
\begin{align}
H_{\mathrm{int}}=&U\sum_{i,\alpha}n_{i\alpha\uparrow}n_{i\alpha\downarrow}+U'\sum_{i,\sigma}\sum_{\alpha<\beta}n_{i\alpha\sigma}n_{i\beta \,-\sigma}\notag\\
   &+(U'-J_H)\sum_{i,\sigma}\sum_{\alpha<\beta}n_{i\alpha\sigma}n_{i\beta\sigma}\notag\\
   &-J_H\sum_{i,\alpha\neq\beta}(c_{i\alpha\uparrow}^{\dagger}c_{i\alpha\downarrow}c_{i\beta\downarrow}^{\dagger}c_{i\beta\uparrow}-c_{i\alpha\uparrow}^{\dagger}c_{i\alpha\downarrow}^{\dagger}c_{i\beta\downarrow}c_{i\beta\uparrow}),
   \label{eq:EQ3}
\end{align}
with intraorbital interaction $U$, interorbital interaction $U'=U-2J_H$ and Hund's coupling $J_H$.

Due to the time dependence in $H_{\mathrm{kin}}(t)$, we apply time
dependent perturbation theory. Here we limit ourselves to the subspace of
states with zero and one doublon-holon (DH) pair, meaning we only
consider virtual $d^5d^5\rightarrow d^6d^4\rightarrow d^5d^5$
excitations \cite{PhysRevB.99.205111}.

Writing down the time-dependent Schr\"odinger equations for this subspace yields 
\begin{align}
 i\delta_t\ket{\Psi_0}_t=&H_{\mathrm{int}}\ket{\Psi_0}_t+\sum_mP_0H_{\mathrm{kin}}(t)\ket{\Psi_{1,m}}_t\label{eq:EQ4}\\
 i\delta_t\ket{\Psi_{1,m}}_t=&P_{1,m}H_{\mathrm{kin}}(t)\ket{\Psi_0}_t+\left(T_{mm}+H_{\mathrm{int}}\right)\ket{\Psi_{1,m}}_t,
 \label{eq:EQ5}
\end{align}
with $P_0$ the projector on the zero DH pair subspace, and $P_{1,m}$
the projector on the one-DH pair subspace with energy $E_m$. 
$\ket{\Psi_{1,m}}$ denotes the subspace of one DH pair with
energy $E_m$.  Propagation of the doublon and
 holon through the lattice within the associated subspace is described by 
 $T_{mm}=P_{1,m}H_{\mathrm{kin}}(t)P_{1,m}$. We dropped $T_{mm'}$ terms, which couple two
energetically distinct subspaces with one doublon holon pair, justified by
\cite{PhysRevB.99.205111}.

$T_{mm}$ can be averaged over time as shown in \cite{PhysRevLett.121.107201}, which yields 
$\bar{T}_{mm}=\mathcal{J}_0(u_{ij})P_{1,m}\hat{v}_{ij}P_{1,m}$.
With this at hand we rewrite (\ref{eq:EQ5}) to
\begin{align}
\ket{\Psi_{1,m}}_t=&\sum_{ij,l}\frac{1}{\Delta E_m-l\omega+\bar{T}_{mm}}e^{-il\omega t}\mathcal{J}_{-l}(u_{ij})v^{m,0}_{ij}\ket{\Psi_0}_t,
\label{eq:EQ6}
\end{align}
with $v^{m,0}_{ij}=P_{1,m}v_{ij}P_0$ and $\Delta E_m=E^1_m-E_0$. Here we performed a partial integration over time and dropped terms $\mathcal{O}(t_h/U^2)$.
Plugging (\ref{eq:EQ6}) into (\ref{eq:EQ4}), we derive the effective time-dependent Hamiltonian
\begin{align}
H_{\mathrm{eff}}(t)=\sum_{ij,k,l,m}v^{0,m}_{ij}\,\frac{\mathcal{J}_{k}(u_{ij})\mathcal{J}_{-l}(u_{ij})e^{i(k-l)\omega t}}{\Delta E_m-l\omega+\bar{T}_{mm}}v^{m,0}_{ij}.
\label{eq:EQ7}
\end{align}
In contrast to the time independent case the
hopping amplitude is now modulated by frequency $\omega$, amplitude
$E_0$ and angle $\varphi$ of the incoming light.

\subsection{Floquet Formalism}\label{floquet}

In Sec.~\ref{eff}, we derived an effective Hamiltonian periodic
in time. It has been shown that Floquet theory can describe these systems with an
effective time-independent Hamiltonian
$\bar{H}_{\mathrm{eff}}$~\cite{PhysRevA.7.2203,PhysRev.138.B979,doi:10.1080/00018732.2015.1055918}. For $\omega\gg t$ a  Magnus
expansion up to first non-vanishing order \cite{KUWAHARA201696} is
sufficient to derive the time independent Floquet Hamiltonian 
\begin{align}
\bar{H}_{\mathrm{eff}}=\sum_{ij,l,m}v^{0,m}_{ij}\frac{\mathcal{J}_{l}(u_{ij})^2}{\Delta E_m-l\omega+\bar{T}_{mm}}v^{m,0}_{ij}.
\label{eq:EQ8}
\end{align}
For $U\gg\lambda\gg t$ the effective Floquet Hamiltonian
can be projected into the $j_{\mathrm{eff}}=1/2$ basis favored by SOC
\cite{PhysRevLett.112.077204,PhysRevLett.102.017205}
\begin{align}
\bar{H}_{\mathrm{eff}}=&\sum_{\braket{ij}\in \alpha\beta(\gamma)}\big[J^{\gamma}\vec{S}_i\vec{S}_j+K^{\gamma}S_i^{\gamma}S_j^{\gamma}+\Gamma^{\gamma}(S_i^{\alpha}S_j^{\beta}+S_i^{\beta}S_j^{\alpha})\notag\\
&+\Gamma'^{\gamma}(S_i^{\alpha}S_j^{\gamma}+S_i^{\gamma}S_j^{\alpha}+S_i^{\beta}S_j^{\gamma}+S_i^{\gamma}S_j^{\beta})\big].
\label{eq:EQ9}
\end{align}
This takes the form of a Kitaev-Heisenberg model, with the
difference being that the factors
$J^{\gamma},K^{\gamma},\Gamma^{\gamma},$ and $\Gamma'^{\gamma}$ 
depend on frequency $\omega$, amplitude $E_0$ and light angle
$\varphi$ in addition to the bond direction $\gamma$. The interaction terms therefore become
\begin{align}
J^{\gamma}=&\sum_{l=-\infty}^{\infty}\mathcal{J}^2_{l}(u_{ij}^{\gamma})\frac{4}{27}\bigg\{\left[-9(t_4^{\gamma})^2+2(t_1^{\gamma}-t_3^{\gamma})^2\right]\notag\\
&\times\mathcal{A}_l(\omega,U,J_H)+(2t_1^{\gamma}+t_3^{\gamma})^2\mathcal{B}_l(\omega,U,J_H)\bigg\}\label{eq:EQ10}\;,\\
K^{\gamma}=&\sum_{l=-\infty}^{\infty}\mathcal{J}^2_{l}(u_{ij}^{\gamma})\frac{4}{9}\left[(t_1^{\gamma}-t_3^{\gamma})^2-3((t_2^{\gamma})^2-(t_4^{\gamma})^2)\right]\notag\\
&\times\mathcal{A}_l(\omega,U,J_H)\;,\label{eq:EQ11}\\
\Gamma^{\gamma} =&\sum_{l=-\infty}^{\infty}\mathcal{J}^2_{l}(u_{ij}^{\gamma})\frac{4}{9}\left[3(t_4^{\gamma})^2+2t_2^{\gamma}(t_1^{\gamma}-t_3^{\gamma})\right]\mathcal{A}_l(\omega,U,J_H)\;,\label{eq:EQ12}\\
\Gamma'^{\gamma} =&-\sum_{n=-\infty}^{\infty}\mathcal{J}^2_{n}(u_{ij}^{\gamma})\frac{4}{9}\left(t_4^{\gamma}(t_1^{\gamma}-t_3^{\gamma}-3t_2^{\gamma}\right)\mathcal{A}_l(\omega,U,J_H)\;,\label{eq:EQ13}
\end{align}
with
\begin{align}
\mathcal{A}_l(\omega,U,J_H)&=g_{dh}(U-3J_H-l\omega)-g_{dh}(U-J_H-l\omega)\;,\notag\\
\mathcal{B}_l(\omega,U,J_H)&=g_{dh}(U+2J_H-l\omega)+2g_{dh}(U-3J_H-l\omega).\label{eq:EQ14}
\end{align}
Notably for LP light the bond dependency additionally enters via
the Bessel functions $\mathcal{J}_l(u_{ij}^{\gamma})$. 
In case of circular polarized (CP) light, the bond
dependence is only influenced by $\mathbf{T}^{\gamma}$ and $u_{ij}$
is bond independent \cite{PhysRevB.103.L100408,Claassen2017}. We will discuss advantages of
both polarization types in more detail in Sec.~\ref{angle}. 

In the pseudospin base the virtual $d^4d^6$ excitations become
$j=1/2,\,j=1/2$ ($d_1$) and $j=3/2,\,j=1/2$ ($d_2$) excitations. 
It has already been shown \cite{PhysRevB.93.214431}, that only the
Heisenberg term (\ref{eq:EQ10}) includes $d_1$
excitations. These are described by the term
$\propto\mathcal{B}_l(\omega,U,J_H)$. Terms
$\propto\mathcal{A}_l(\omega,U,J_H)$ in
(\ref{eq:EQ10})-(\ref{eq:EQ13}) describe a $d_2$
excitation.  

The Greens function $g_{dh}$ captures the propagation of the DH pair before decaying 
into the $d^5d^5$ subspace again 
\begin{align}
g_{dh}=&\bra{\Psi_0}c_{j\beta\sigma}^{\dagger}c_{i\alpha\sigma}
\frac{1}{\Delta E_m-l\omega+\bar{T}_{mm}}P_{1,m}c_{i\alpha\sigma}^{\dagger}c_{j\beta\sigma}\ket{\Psi_0}.
\label{eq:EQ15}
\end{align}
Since we neglected $T_{mm'}$, the DH propagation takes only place in the subspace 
with energy $E_m$. 

If $|\Delta E_m-l\omega|\gg \bar{T}_{mm}$, i.e. for off-resonant
incoming light, the propagation in the one DH subspace is negligible,
meaning the Green's function simply becomes 
$g_{dh}(\Delta E_m-l\omega)\approx 1/(\Delta E_m-l\omega)$. Heating effects
near the resonance are neglected in this off-resonance approximation (ORA). 

In case of $\Delta E_m\approx l\omega$, the propagation of the DH pair in the excited subspace
$T_{mm}$ is no longer negligible,
and the ORA may break down. We therefore
investigate the retracable path approximation (RPA) valid near resonances. 
The RPA \cite{PhysRevB.2.1324}
is a method to capture the behavior near the resonance for a one-band
model. This method was already used to capture heating in multi-band spin-orbital
models for $\mathrm{YTiO}_3$ and $\mathrm{LaTiO}_3$
\cite{PhysRevB.99.205111,PhysRevLett.121.107201}.    

Let us first comment on the RPA's applicability. If there was only the
$j=\tfrac{1}{2}$ doublet, the Kitaev-Heisenberg model could be seen as
a quasi one-band model. In reality, however, there
is also the possibility of $d_2$
in addition to $d_1$ doublons, i.e., excitation of a hole into the $j=\tfrac{3}{2}$ states. 
Additionally, the
four $j=\tfrac{3}{2}$ states correspond to two 'bands' already by
themselves. A one-band approach can then
only work as long as mixing between $d_1$ and
$d_2$ doublons remains small and if the $j=\tfrac{3}{2}$ states can be neatly separated into two
weakly mixed bands. Fortunately,  resonant inelastic X-ray experiments indicate that this is the
case, because trigonal distortion splits the two $j=\tfrac{3}{2}$ sub-bands~\cite{PhysRevLett.110.076402,Lebert_2020,Yadav2016}, which
suppresses  their mixing. The Green's functions relating to
$d_1$  doublons and to the two sub-bands can then be
approximated as independent one-band problems. Projecting (\ref{eq:EQ15}) into the
pseudospin basis leaves us then with different Green's functions, for
$d_1$  doublons with hopping amplitudes $t_{B,1}$ and for
$d_2$ doublons with  hopping
amplitude $t_{B,2}$. 

In order to perform the RPA we first rewrite the DH Green's function in
terms of a holon $g_{h}$ and a doublon $g_{d}$ Green's function,
assuming that holon and doublon movements are uncorrelated. This can
be done by a convolution integral as shown in
\cite{PhysRevB.99.205111,PhysRevLett.121.107201}. Only doublon and holon paths in the excited subspace that end at
their starting  point are allowed. In addition, we
exclude "loop" paths,  leaving only retracable paths \cite{PhysRevB.2.1324}. 
A detailed derivation can be found in \cite{PhysRevB.99.205111} 
 
For this approximation the Greens function becomes complex and 
no longer diverge at $\Delta E_m=l\omega$. Its imaginary part
describes the heating due to DH propagation. 
The bandwidth of the heating is
$2\sqrt{z-1}t_B$ \cite{PhysRevLett.121.107201}, with the coordination number of the lattice $z$ and the DH hopping strength $t_B$. In the remainder of the paper we set $t_B=t_{B,1}=t_{B,2}$.

\section{Results}\label{res}

\subsection{Materials}\label{mat}
\begin{table}
\begin{tabularx}{\columnwidth}{XXXXXXX}
\toprule
Material&\multicolumn{2}{c}{$\mathrm{Na}_2\mathrm{IrO}_3$}&\multicolumn{2}{c}{$\alpha$-$\mathrm{Li}_2\mathrm{IrO}_3$} & \multicolumn{2}{c}{$\alpha$-$\mathrm{RuCl}_3$}\\
\midrule
$U$&\multicolumn{2}{c}{$1.7$}&\multicolumn{2}{c}{$1.7$}&\multicolumn{2}{c}{$3.0$}\\
$J_H$&\multicolumn{2}{c}{$0.3$}&\multicolumn{2}{c}{$0.3$}&\multicolumn{2}{c}{$0.5$}\\
\midrule
NN&1\textsuperscript{st}&3\textsuperscript{rd}&1\textsuperscript{st}&3\textsuperscript{rd}&1\textsuperscript{st}&3\textsuperscript{rd}\\
\midrule
$t_1^{z}$&33.1&-9.3&55.0&-6.4&50.9&-8.2\\
$t_2^{z}$&264.3&-13.8&219.0&-13.5&158.2&-7.4\\
$t_3^{z}$&26.6&-36.8&-175.1&-33.3&-154.0&-39.5\\
$t_4^{z}$&-11.8&16.6&-124.5&16.6&-20.2&11.7\\
$t_1^{x/y}$&38.8&-8.4&76.3&-6.3&45.4&-7.7\\
$t_2^{x/y}$&269.3&-12.7&252.7&-13.4&162.2&-7.8\\
$t_3^{x/y}$&-19.4&-35.3&-108.8&-33.0&-103.1&-41.4\\
$t_4^{x/y}$&-23.4&16.0&-9.3&15.8&-13.0&11.7\\

\bottomrule
\end{tabularx}
\caption{Parameters for $\mathrm{Na}_2\mathrm{IrO}_3$,
  $\alpha$-$\mathrm{Li}_2\mathrm{IrO}_3$, and $\alpha$
  -$\mathrm{RuCL}_3$. 1\textsuperscript{st}- and 3\textsuperscript{rd}-nearest--neighbor (NN) hopping parameters in meV are taken from
  \cite{PhysRevB.93.214431}. Coulomb repulsion $U$  
and Hund's coupling in eV taken from \cite{PhysRevLett.113.107201} for
iridates and from \cite{PhysRevB.91.241110} for
ruthenates} \label{tab:Tab1} 
\end{table}
As promising candidates for a realization of the $J$-$K$-$\Gamma$
model several iridates and ruthenates have been suggested
\cite{PhysRevB.88.035107,PhysRevB.91.241110,PhysRevLett.109.197201,PhysRevLett.112.077204,PhysRevB.93.214431,Winter_2017}. In
this paper we will focus on the three compounds
$\mathrm{Na}_2\mathrm{IrO}_3$, $\alpha$-$\mathrm{Li}_2\mathrm{IrO}_3$,
 and $\alpha$-$\mathrm{RuCl}_3$. In the following discussion we use the
hopping parameters of \cite{PhysRevB.93.214431} with the only
distinction that we chose symmetrized hopping parameters $t_1^{x/y}=(t_{1a}'+t_{1b}')/2$ and
$t_4^{x/y}=(t_{4a}'+t_{4b}')/2$. The values of Hund's coupling $J_H$
and Coulomb repulsion $U$ for iridates and ruthenates  are taken from
\cite{PhysRevLett.113.107201} and \cite{PhysRevB.91.241110}
respectively. All considered parameters are listed in
Tab.~\ref{tab:Tab1}.  

Notably all materials possess an anisotropy in $z$-direction, which
intrinsically calls for bond dependent treatment of $J$,$K$, and
$\Gamma$.

\subsection{Heating in dependency of $\omega$ and $E_0$}\label{heat}
\begin{figure}
\includegraphics[width=\textwidth]{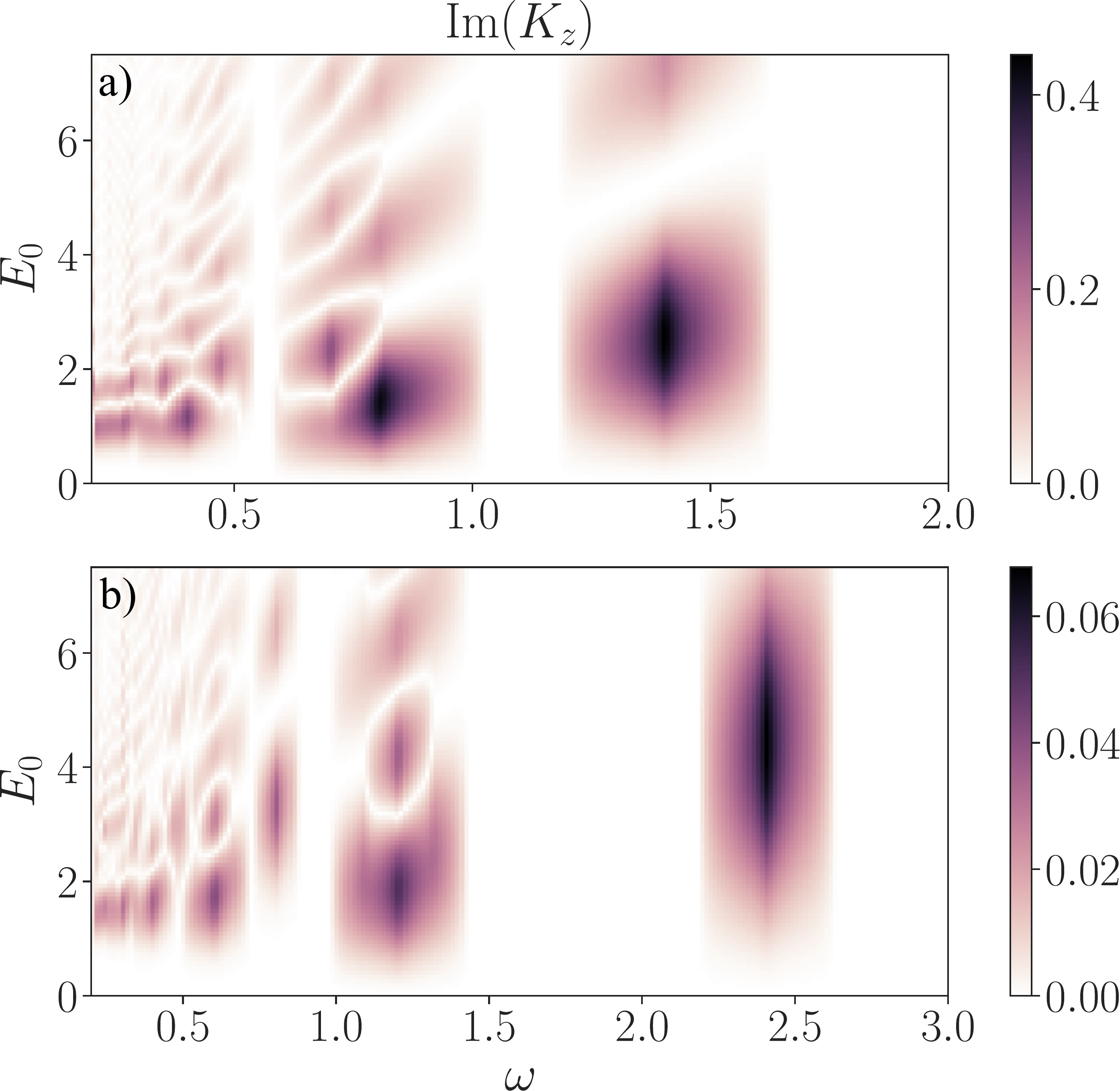}
\caption{Imaginary part of $K_z$ plotted for varying light frequency $\omega$ and amplitude $E_0$ in eV, with the parameter set of $\text{Na}_2\text{IrO}_3$ a) and $\alpha$-$\text{RuCl}_3$ b).
For the calculations of the Green's functions the retracable path approximation (RPA) was used.}
\label{fig:Fig1}
\end{figure}
In this section we investigate possible heating effects arising from DH propagation captured by the RPA. The bandwidth $t_B$ is set to $t_B=0.04\,$eV.
Representative for the other interactions, Kitaev interactions are analyzed in dependency of the frequency $\omega$ and the amplitude $E_0$ of the LP light.
We only discuss z-Bond terms to get an qualitative understanding of heating effects.
To do so we examine the imaginary part of the Kitaev interaction $K_z$ (\ref{eq:EQ11}) calculated with RPA.  

For $\text{Na}_2\text{IrO}_3$ [Fig.~\ref{fig:Fig1}(b)] heating is present around a relatively broad
area around the resonances, leaving a narrow area around $\omega=1.1\,\text{eV}$, where 
heating-free driving is possible.
In this area, from now on referred to as driving corridor the ORA is expected to be valid. 
For frequencies $\omega<U-3J_H$ driving without heating becomes difficult due to a multitude
of resonances.
Evidently the width of the driving corridor is not only dependent on $t_B$ but also the magnitude
of $J_H$. Increasing $J_H$ separates the resonances and therefore enhances the
driving corridor. A broad driving corridor is a desirable feature for experiments.
In $\text{Li}_2\text{IrO}_3$ heating effects are very similar to $\mathrm{Na}_2\mathrm{IrO}_3$.
Therefore we omit a more detailed discussion in this section.   

In comparison to iridates, $\alpha$-$\text{RuCl}_3$ has significant larger Coulomb repulsion and
Hund's interaction. Fig.~\ref{fig:Fig1}(b) clearly displays that the resonance peaks are
further apart than in Fig.~\ref{fig:Fig1}(a), causing a broader driving corridor.
This makes $\alpha$-$\text{RuCl}_3$ a more promising candidate for Floquet
engineering than iridate compounds. Also the ORA appears to be well
justified within the driving corridor. 

We conclude that both iridates and ruthenates are susceptible to Floquet engineering within a
certain driving corridor. This driving corridor is mainly determined by $J_H$ and $t_{B}$.
Materials with large Hund's interaction and small $t_B$ should be primed for
Floquet engineering.
Within this driving corridor the ORA is expected to be valid and differences to the RPA should be  minute.

\subsection{Floquet engineering in the ORA}\label{eng}
\begin{figure}
\includegraphics[width=\textwidth]{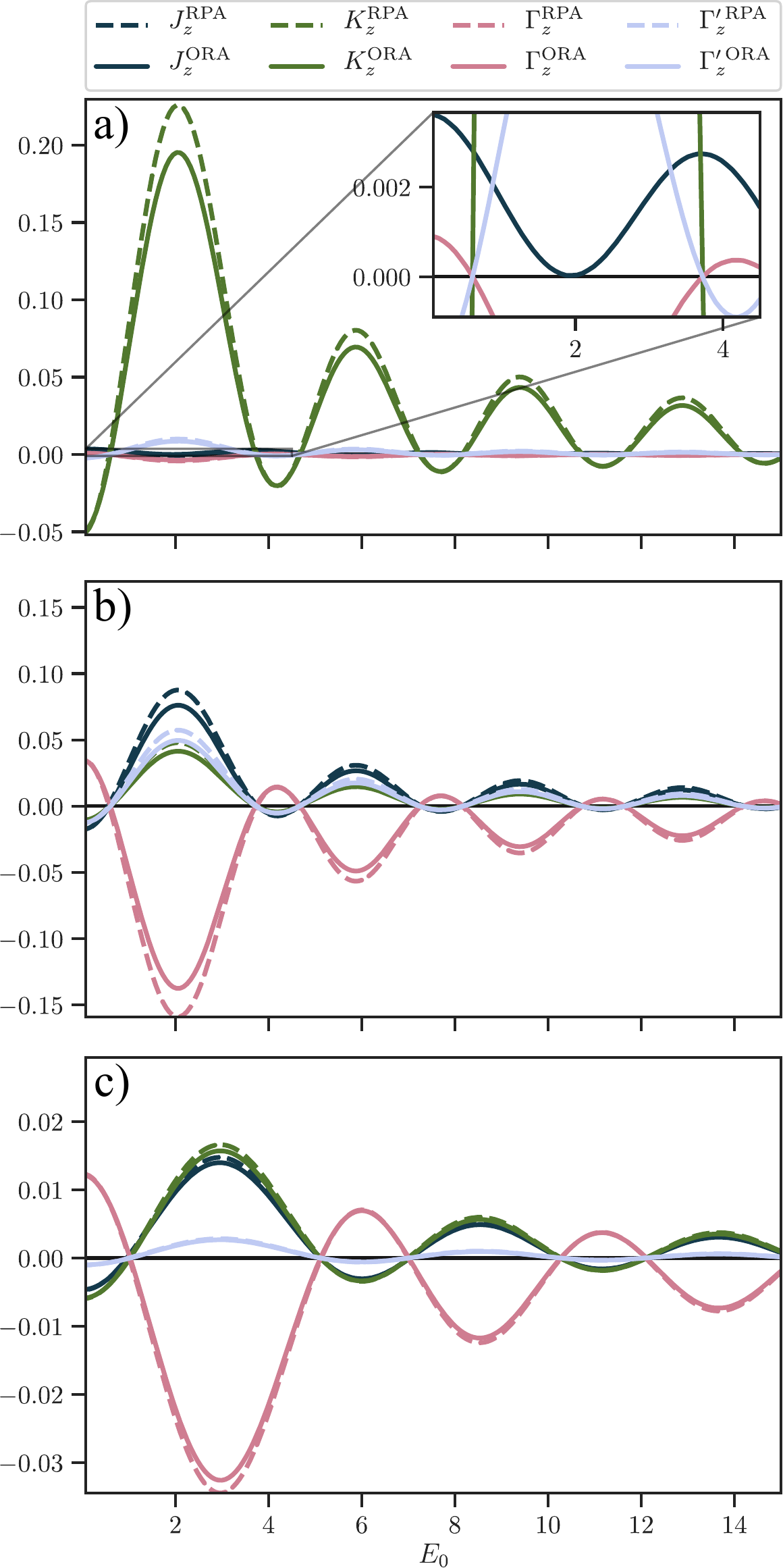}
\caption{$J$, $K$, $\Gamma$, and $\Gamma'$ terms in dependency of the light amplitude $E_0$ in eV for $\mathrm{Na}_2\mathrm{IrO}_3$ parameter set with fixed frequency $\omega=1.1\,$eV a). $\alpha$-$\mathrm{Li}_2\mathrm{IrO}_3$ with $\omega=1.1\,$eV, and $\alpha$-$\mathrm{RuCl}_3$ with $\omega=1.6\,$eV are displayed in b) and c) respectively.  The real part of the RPA is displayed with dashed lines, the ORA with solid lines. The inset in a) shows a zoom in for $0<E_0<5$.}
\label{fig:Fig2}
\end{figure} 
After considering possible influences of heating, we focus on
 Floquet driving within the driving corridor between
the resonances at $U-3J_h$ and $U-J_H$. To do so we set the
frequency to $\omega=1.1\,$eV for iridates and $\omega=1.6\,$eV for
the ruthenate. As we can see in Fig.~\ref{fig:Fig2}, varying $E_0$ can change
both magnitude and sign of the interaction terms. In addition to ORA
results (solid lines), the results for the RPA with $t_B=0.4$ are displayed (dashed lines). 
Since the differences are minor and heating is negligible (see Sec.~\ref{heat}), 
we will from here on
only discuss the ORA results.  

The interaction terms for $\mathrm{Na}_2\mathrm{IrO}_3$ 
are shown in Fig.~\ref{fig:Fig2}(a). The
Kitaev interaction is dominant throughout the whole parameter range,
while the other terms are considerably smaller. We also note that the
absolute magnitude of the interactions decreases
with increasing $E_0$. 
The inset in Fig.~\ref{fig:Fig2}(a) clearly displays that $J$ can be tuned
relative to the $K$, $\Gamma$ and $\Gamma'$ interactions, by changing $E_0$.

On the z-bond of $\mathrm{Na}_2\mathrm{IrO}_3$ this
effect is particularly visible since the factor $2t_1+t_3$ in
$\mathcal{B}_l(\omega,U,J_H)$ becomes relatively large, due to $t_1>0$
and $t_3>0$. This effect has already been discussed in
\cite{PhysRevB.103.L100408}. However this has been done with a change of sign in $t_3$ for
$\alpha$-$\mathrm{RuCl}_3$. It is encouraging that this
effect also arises for realistic parameters in
$\mathrm{Na}_2\mathrm{IrO}_3$. For a tuning of $J$ it is
sufficient to have a comparable energy scale of $\mathcal{A}_l(\omega,U,J_H)$ and $\mathcal{B}_l(\omega,U,J_H)$.  On the $z$-Bond of $\mathrm{Na}_2\mathrm{IrO}_3$, we therefore have the
ability to completely turn off the Heisenberg interactions at
$E_0\approx2\,$eV and reduce the model to a
$K$-$\Gamma$-model. However this is not the case for the x- and y-Bond
in $\mathrm{Na}_2\mathrm{IrO}_3$, due to the fact that
$t_3^{x/y}<0$ for these bonds, which reduces the influence of
$\mathcal{B}_l(\omega,U,J_H)$. 

In $\mathrm{Li}_2\mathrm{IrO}_3$ the Kitaev interaction is far less
dominant, and the $J$, $\Gamma$, $\Gamma^{\prime}$ terms all are significant [Fig.~\ref{fig:Fig2}(b)]. 
Moreover, the values of $t_1$ and $t_3$ (see Tab.\ref{tab:Tab1}) imply a low tunability of $J$. In Fig.~\ref{fig:Fig2}(b) we indeed see that the
$J$ term is almost in phase with the other terms.  
It is however still possible to adjust magnitude and sign of
the interaction terms with the help of light frequency and amplitude.  
     
For $\alpha$-$\mathrm{RuCl}_3$ results
for all interaction parameters are shown in Fig.~\ref{fig:Fig2}(c). As
in the case of $\alpha$-$\mathrm{Li}_2\mathrm{RuO}_3$, the $J$
parameter is mostly in phase with the other terms. Again the
reason for that is the sub-optimal combination of the $t_1$ and $t_3$
parameters. Compared to iridates, the frequency of
the oscillation is significantly lower. The origin is the larger
$\omega$ needed to suppress heating in ruthenates, which enters the 
Bessel function as $\frac{1}{\omega}$.  

The results of Fig.~\ref{fig:Fig2}(c) thus confirm \cite{PhysRevB.103.L100408}, in that
for realistic parameters and only considering NN
interactions a tuning of $J$, $K$, $\Gamma$, and $\Gamma'$ relative to
each other is not possible. However we found that in other materials
like $\mathrm{Na}_2\mathrm{IrO}_3$ a tuning of the Heisenberg term is
possible due to a more suitable combination of the
hopping parameters.

\subsection{Third-nearest--neighbor Heisenberg interactions}\label{tnn}

\begin{figure}
\includegraphics[width=\textwidth]{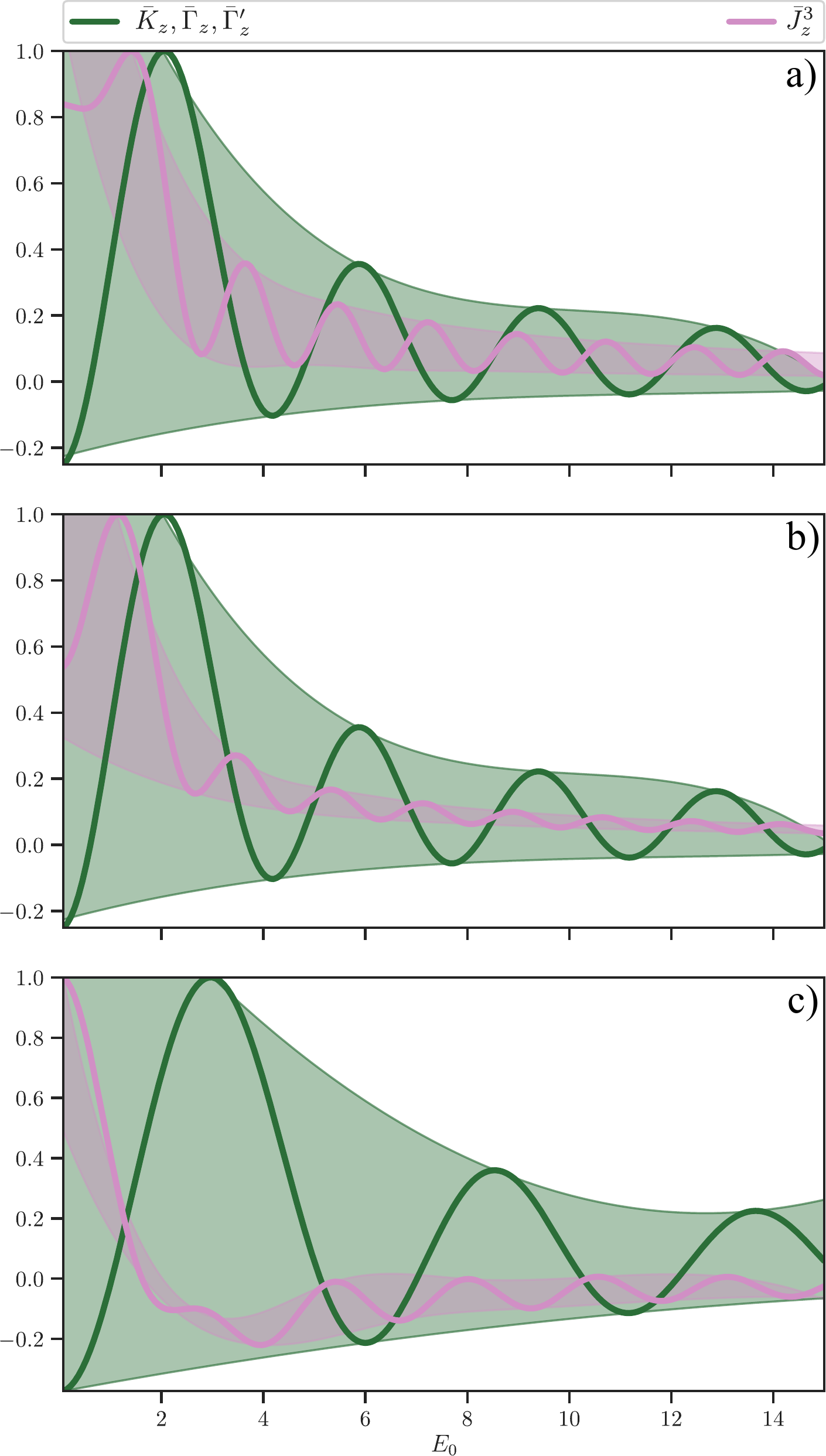}
\caption{Comparison of normalized $\bar{K}_z, \bar{\Gamma}_z,\bar{\Gamma}^{\prime}$ and third-nearest--neighbor Heisenberg $\bar{J}_z^3$ interactions in dependency of $E_0$ in eV. For $\mathrm{Na}_2\mathrm{IrO}_3$ a) and $\alpha-\mathrm{Li}_2\mathrm{IrO}_3$ b) the frequency is set to $\omega=1.1\,\mathrm{eV}$. For $\alpha\mathrm{RuCl}_3$ c) the frequency is $\omega=1.6\,\mathrm{eV}$.    }
\label{fig:Fig3}
\end{figure}

Third-nearest--neighbor (3\textsuperscript{rd}$\,$NN) Heisenberg interactions are argued to have  a
significant impact on the formation of the zig-zag ground state
found in iridates and
ruthenates \cite{PhysRevB.96.064430}. Especially in $\mathrm{Na}_2\mathrm{IrO}_3$, $J_3$ is a
main reason for the absence of the KSL \cite{PhysRevB.93.214431}. For
this reason, we investigate the possibility of tuning  
$J_3(E_0,\omega)$ relative to the NN interactions.

$J_3$ is calculated in the same manner as $J$ [with (\ref{eq:EQ10})]. The
hopping parameters considered for $J_3$ are taken from
\cite{PhysRevB.93.214431} and listed in Tab.~\ref{tab:Tab1}.  
We note that the requirement for a significant tuning, a large
$\left|(2t_1+t_3)\right|$, is fulfilled for all materials. In
addition, the larger distance $|r_{ij}|$ for third nearest
neighbors changes the frequency of $\mathcal{J}_{-l}(u_{ij})$. Together
with the promising hopping parameters, this suggests that $J_3$ can be
tuned relative to the other interaction parameters, similar to  $J_z$ in $\mathrm{Na}_2\mathrm{IrO}_3$. 

Since we want to compare interaction terms of bonds with different directions (3\textsuperscript{rd}$\,$NN and NN) CP light is more suitable. This is the case since for CP $u_{ij}^{c}=e/\omega E_0|\mathbf{r}^{\gamma}_{ij}|$
becomes bond independent, due to $|\mathbf{r}^{\gamma}_{ij}|=|r_{ij}|$
for NN's. The only distinctions between NN and 3\textsuperscript{rd}$\,$NN are therefore the distance $|r_{ij}|$ and the hopping parameters. 
  
In Fig.~\ref{fig:Fig3} the normalized $\bar{K}$, $\bar{\Gamma}$, $\bar{\Gamma}^{\prime}$ and $\bar{J}_3$ interactions are shown. As normalization factor the maximum of the respective interaction in the considered $E_0$ range is chosen. It becomes obvious that for
all materials $J^{3}$ decreases significantly faster with $E_0$ than
the NN interactions. This gives a pathway to suppress $J^{3}$ via the
light amplitude $E_0$. Due to the differing frequencies for NN and 3\textsuperscript{rd}$\,$NN
it is also possible to completely turn off $J^{3}$ while having finite
NN interactions. 

In general CP light is well suited for isotropic materials since it
does not change the magnitude of bond interactions relative to each
other. However if a material has anisotropic interaction terms, LP
light has significant advantages over CP light, which we discuss in
the following section.

\subsection{Light Angle}\label{angle}

\begin{figure}
\includegraphics[width=\columnwidth]{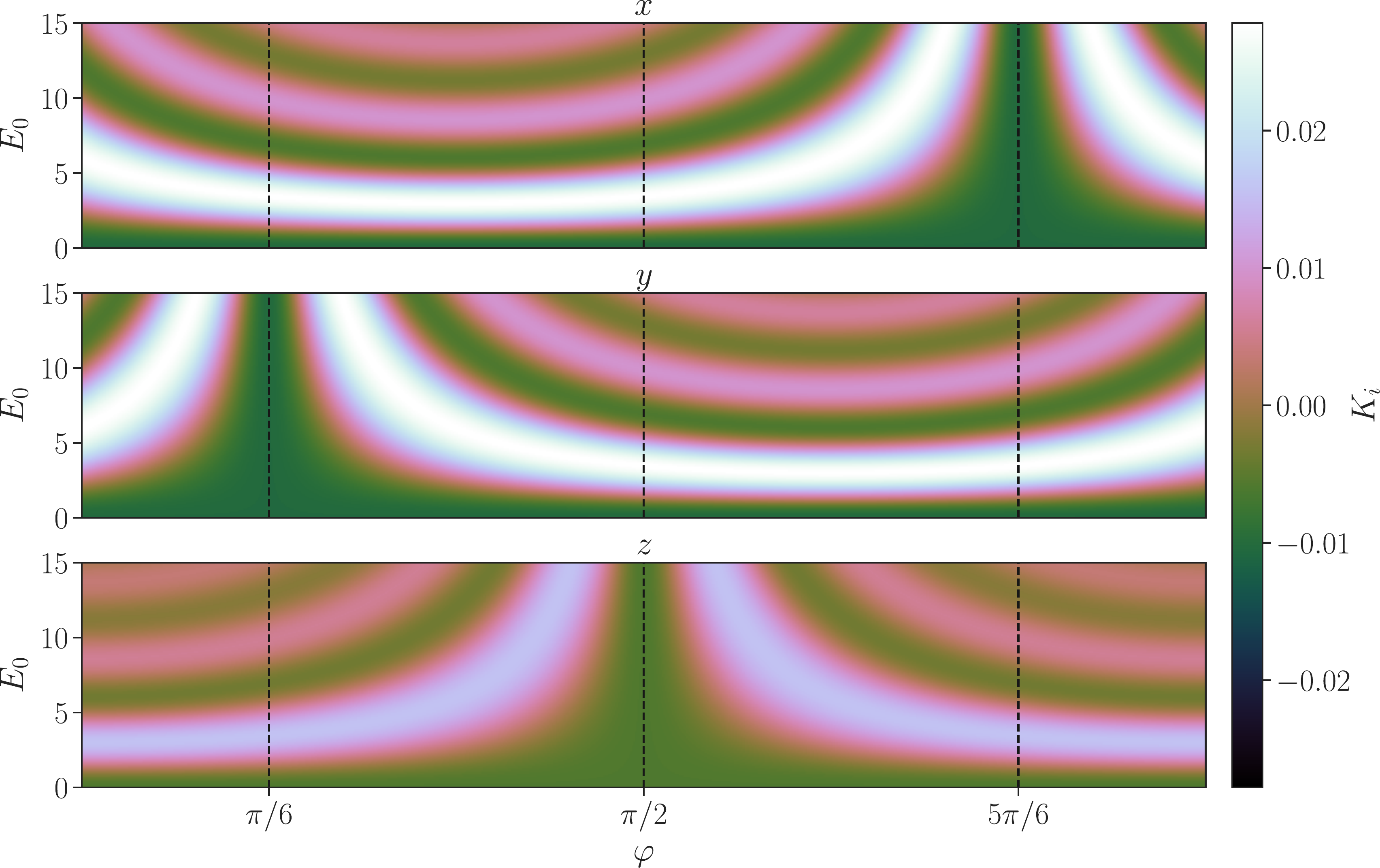}
\caption{The magnitude of the Kitaev interaction term for $x$-, $y$- and $z$-bond depending on the light angle $\varphi$ and the light amplitude $E_0$ in eV is shown for $\alpha$-$\mathrm{RuCl}_3$.}
\label{fig:Fig4}
\end{figure} 
\begin{figure*}
\includegraphics[width=\textwidth]{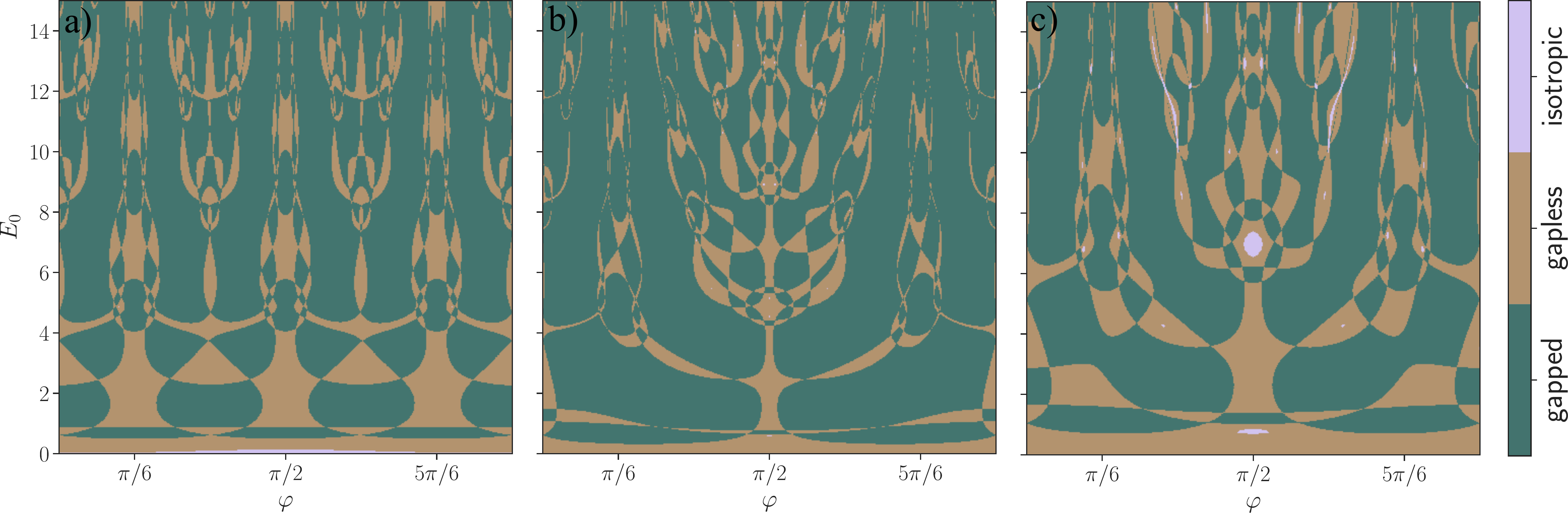}
\caption{$E_0$-$\varphi$ phase diagram where Kitaev interactions of
  Fig.~\ref{fig:Fig4} would give rise to a gapped (light blue), gapless
  (purple) and isotropic (orange) Kitaev phase. The terms gapless and
  isotropic are defined by the criteria in the text. Calculations
  were done for $\mathrm{Na}_2\mathrm{IrO}_3$ a),
  $\alpha$-$\mathrm{Li}_2\mathrm{IrO}_3$ b), and
  $\alpha$-$\mathrm{RuCl}_3$ c).} 
\label{fig:Fig5}
\end{figure*}

Up until now we only considered LP light parallel to the
$z$-axis. In general LP light can point in an arbitrary 
direction, making it a parameter that is experimentally adjustable. The angle of the incoming light has an impact
on the interaction terms, therefore it should be possible
to tune the interactions by just changing the light angle $\varphi$. 
We expect the light angle $\varphi$ to alter the relative strengths of
the $x$, $y$ and $z$ interaction terms, hence we now consider the
$x$ and $y$ bond in addition to the $z$ bond.

The focus in this
chapter is tuning the Kitaev interactions, which is why for the
remainder of the section we only consider $K_x$, $K_y$, and $K_z$. 
We calculate the Kitaev interactions depending on the
light angle $\varphi$ and the driving amplitude $E_0$ with the ORA. Since the
interaction terms are $\pi$-periodic we only consider $\varphi\in
[0,\pi ]$. The results for $\alpha$-$\mathrm{RuCl}_3$ at $\omega=1.6\,$eV
are displayed in Fig.~\ref{fig:Fig4}. 

The corridors at $\varphi=\pi/6$, $\pi/2$, and $5\pi/6$ for the $y$-,
$z$- and $x$-bond interaction respectively are the most notable
features. For these angles the Kitaev term of the respective bond is
independent of $E_0$. We can see that if $\mathbf{r}_{ij}\perp
\mathbf{A}(t)$ the interaction decouples from the light, since $u_{ij}=0$. Therefore it
is possible to tune two bond interactions while the coupling along the
remaining bond direction remains unchanged. This opens the path to eliminate
anisotropies intrinsic to, e.g.,
$\alpha$-$\mathrm{Li}_2\mathrm{IrO}_3$ and
$\alpha$-$\mathrm{RuCl}_3$.  

The second effect we observe is that depending on the light angle and
the driving amplitude, the Kitaev interactions can have different signs
for different bonds. Additionally, the magnitude of the
interactions also drastically depends on the light angle $\varphi$. As
a consequence of these two effects, the Kitaev interactions might no
longer support a gapless Kitaev phase, but rather give rise to a
gapped Kitaev phase, due to the strong anisotropies.

We are aware that in all considered materials the Kitaev phase is not yet realized. In this section we therefore want to discuss the nature of the Kitaev phase if it were to be realized. Interactions giving rise to a gapped (gapless) Kitaev phase are referred to as gapped (gapless) interactions.

The criterion for
gapless interactions $K_{\alpha}+K_{\beta}>K_{\gamma}$, with
$\alpha,\beta,\gamma\in (x,y,z)$, does not hold for arbitrary
$\varphi$ and $E_0$. Applying this condition to the Kitaev terms of
Fig.~\ref{fig:Fig4}, we see where gapped and gapless interactions arise. The results for the different materials are
displayed in Fig.~\ref{fig:Fig5}. We also added  
 isotropic interactions with the condition $|K_{\alpha}-K_{\beta}|<\epsilon$,
for all combinations of $\alpha,\gamma$ and $\epsilon=10^{-3}$.
This isotropic interaction is included to highlight that Floquet
engineering is capable of increasing the isotropy in materials
[Fig.~\ref{fig:Fig5}(c)]. 

For $\mathrm{Na}_2\mathrm{IrO}_3$ we observe that without driving the
NN Kitaev interactions are already almost
isotropic. Turning on the LP light then breaks the isotropy and also
gives rise to gapped interactions. Around $\varphi=\pi/6$, $\pi/2$, and
$5\pi/6$ the interactions are very likely to be gapless.  

In case of $\alpha$-$\mathrm{Li}_2\mathrm{RuO}_3$ the system has
gapless interactions in the absence of LP light. Again turning on LP
light can lead gapped interactions. In contrast to
$\mathrm{Na}_2\mathrm{IrO}_3$, there is only a preference of  gapless interactions at $\pi/2$. This can be explained due to the large
intrinsic anisotropy in $z$-direction. Also the areas of a gapped
interactions are significantly larger than in
$\mathrm{Na}_2\mathrm{IrO}_3$ leading to the conclusion that deriving
a gapless interaction in the presence of a LP light field might be
complicated. 

Like in $\alpha$-$\mathrm{Li}_2\mathrm{RuO}_3$,
the high intrinsic anisotropy in $\alpha$-$\mathrm{RuCl}_3$ allows a stable gapless interaction only at $\pi/2$. However
unlike the case in the iridate, the ruthenate actually can be driven
into a isotropic interaction at $\varphi\approx \pi /2$ and
$E_0\approx 7\,\text{eV}$.

\section{Disussion and Conclusions}\label{summary}

We have investigated Floquet engineering for a variety of
candidate Kitaev materials. 
To do so we derived an effective Floquet Kitaev-Heisenberg model for LP light
similar to the CP light model of \cite{PhysRevLett.112.077204}. The Green's functions in
(\ref{eq:EQ15}) were approximated with the ORA and RPA.  

With help of the RPA we were able to determine a frequency range
where the ORA is valid. In addition, we also
identified frequencies where heating can be assumed to be negligible. 
This lead us to the conclusion that Floquet engineering in ruthenates
should in principle be more promising than in iridates.   

For suitable off-resonance frequencies, we investigated the influence of the
light amplitude on the nearest--neighbor interaction terms (Fig.~\ref{fig:Fig2}),
where our results qualitatively agree with the results reported in
Ref.~\cite{PhysRevB.103.L100408}. We thus  
observe that tuning in magnitude and sign of the interactions in
dependency of the light amplitude is possible.  For instance, Heisenberg $J$ can be
tuned  relative to the other parameters, especially on the $z$-bond of
the $\mathrm{Na}_2\mathrm{IrO}_3$ compound.   

Going beyond nearest--neighbors, we find that third-nearest--neighbor Heisenberg interactions $J_3$
\cite{PhysRevB.93.214431} are quite susceptible to Floquet tuning. With the right setting of frequency and amplitude, a
significant decrease of the third-nearest--neighbor interaction terms
should be achievable in all considered materials. Circular polarized
is expected to work better than linear polarization, and could be used
to suppressing zig-zag order, which is stabilized by $J_3$ and  competes with the Kitaev spin liquid.

Linear polarized light, on the other hand, allows us  to tune the interactions in different directions, with the
help of the angle of the incoming light. Starting from a pure Kitaev
model, this can either lead to a gapped model or more isotropic Kitaev
interactions. It is also possible to decouple single bonds completely
from the light so that they remain unaffected, so one can specifically
"erase" anisotropies in one direction. 

After investigating the more traditional Kitaev candidate materials,
it would be interesting to consider materials which at
first glance seem unattractive due to e.g. high anisotropies, which
could be tuned into promising candidates with the help of Floquet
engineering. Finally, materials with strong $U$ and $J_H$ are particularly 
promising, due to a wide driving corridor. This
suggests materials based on $3d$
elements~\cite{PhysRevLett.125.047201,PhysRevB.102.224429}, where these interactions tend to be stronger.

%%%%%%%%%%%%%%%%%%%%%%%%%%%%%%%%%%%%%%%%%%%%%%%
\bibliography{References}

\end{document}